%Paper: hep-th/9410242
%From: sonoda@physics.ucla.edu (Sonoda)
%Date: Mon, 31 Oct 94 18:14:41 -0800
%Date (revised): Tue, 1 Nov 94 11:19:23 -0800

% October 1994
\input harvmac
\noblackbox
\def\L{\Lambda_{\rm QCD}}

\Title{UCLA/94/TEP/41}{A Scheme Independent Definition of $\L$$^\star$}
\footnote{}{$^\star$ This work was supported in part
by the U.S. Department
of Energy, under Contract DE-AT03-88ER 40384 Mod A006 Task C.}

\centerline{Hidenori SONODA$^\dagger$\footnote{}{$^\dagger$
sonoda@physics.ucla.edu}}
\bigskip\centerline{\it Department of Physics,
UCLA, Los Angeles, CA 90024-1547, USA}

\vskip 1in
Given a renormalization scheme of QCD, one can define
a mass scale $\L$ in terms of the beta function.
Under a change of the renormalization scheme,
however,
$\L$ changes by a multiplicative constant.
We introduce a scheme independent $\L$
using a connection on the space of the coupling
constant.

\def\dt{{d \over dt}~}
\def\e{{\rm e}}
\def\tg{\tilde{g}}
\def\tx{\tilde{x}}
\def\tL{\tilde{\Lambda}_{\rm QCD}}
\def\O{{\cal O}}
\def\Lag{{\cal L}}
\def\cg{(c_g)_g^{~g}}
\def\ctg{(c_{\tg})_{\tg}^{~\tg}}

%\draft
\Date{October 1994}

QCD with no quarks is parametrized only by
the strong fine structure constant $g \equiv \alpha_s$.  There is no
physical meaning to the value of the parameter $g$
for two reasons.  First, the value of $g$ changes
as we change the scale at which we define the theory.
This scale dependence is given by the renormalization
group (RG) equation:
\eqn\eRGg{\dt g = \beta^g (g) ,}
where the beta function can be expanded as
\eqn\ebeta{\beta^g (g) = {1 \over 2} b_1 g^2 + {1 \over 3!} b_2 g^3
+ O(g^4) .}
Second, the parameter changes when we change the renormalization
scheme.  Changing a scheme is equivalent to redefining $g$ by
\eqn\escheme{\tilde{g} = f(g) ,}
where $f(g)$ is an arbitrary regular function constrained by
\eqn\ederivative{f'(0) = 1 .}
For $\tg$, the RG equation is given by
\eqn\ebetatilde{\dt \tg = \beta^{\tg} (\tg) ,}
where
\eqn\edefbetatilde{\beta^{\tg} (\tg) \equiv f'(g) \beta^g (g) .}
We use the convention in which the RG runs toward the infrared.
We will measure quantities of mass dimension
$d$ in units of $\mu^d$, where $\mu$ is the renormalization
scale.  Hence, we can take $\mu =1$ from now on.

The function $\L (g)$ gives a physical mass scale,
and therefore it must satisfy the RG equation
\eqn\eRG{\dt \L (g) \equiv \beta^g (g) {d \over dg} \L (g) = \L (g) .}
The solution, up to a multiplicative constant, is given by
\ref\rcollins{See, for example, J.~Collins, {\it Renormalization}
(Cambridge University Press, 1984).}
\eqn\elambda{
\L (g) = \e^{- {2 \over b_1 g}} g^{- {2 b_2\over 3 b_1^2}}
\exp \left[
\int_0^g dx \left( {1 \over \beta^g (x)} - {2 \over b_1 x^2}
+ {2 b_2 \over 3 b_1^2 x} \right) \right] ,}
where $b_1, b_2$ are defined in Eqn.~\ebeta.
Unfortunately, this definition depends on the scheme.
\ref\rstirling{See, for example,
P.~H.~Frampton,
{\it Gauge Field Theories} (Benjamin/Cummings,
1987); R.~K.~Ellis and W.~J.~Stirling,
``QCD and Collider Physics,'' lectures at
1988 CERN School of Physics.}

To see the scheme dependence, we compare $\L(g)$, given by \elambda,
with $\tL (\tg)$ for $\tilde{g}$, defined analogously by
\eqn\elambdatilde{
\tL (\tg)
= \e^{- {2 \over b_1 \tg}} \tg^{- {2 b_2\over 3 b_1^2}}
\exp \left[
\int_0^{\tg} d\tx \left( {1 \over \beta^{\tg} (\tx)} - {2 \over b_1 \tx^2}
+ {2 b_2 \over 3 b_1^2 \tx} \right) \right] .}
(Recall the scheme independence of the coefficients $b_1, b_2$.)
Since the ratio $\tL (\tg)/\L(g)$ is invariant under the RG, it is a
constant.  Hence, we can evaluate the ratio at the origin $g=0$.
We find
\eqn\eratio{{\tL (\tg) \over \L (g)} = \e^{a \over b_1} ,}
where the constant $a$ is given by
\eqn\ea{a \equiv f''(0) .}

To remove the scheme dependence, it is not enough to
know the beta function.  We need an additional quantity
whose scheme dependence cancels
that of the beta function.
In ref.~\ref\rsonoda{H.~Sonoda, ``Connection on the theory space,''
hep-th/9306119} a connection on the theory space has been
introduced.  In this particular case the theory space is simply a
line with a coordinate $g$.  The change of $g$ is
generated by a conjugate field $\O_g$.  If we write
the lagrangian density as
\eqn\elagrangian{\Lag = {1 \over 4g} F^2 ,}
then the conjugate field is given by
\eqn\econjugate{\O_g \equiv {\partial \Lag \over \partial g}
= - {1 \over 4 g^2} F^2 .}
(The above expression can be made more precise.
But it is not important here.)
Under the coordinate change \escheme, the conjugate field
transforms like a tangent vector ${d \over dg}$:
\eqn\etransO{\O_{\tg} = {dg \over d\tg} \O_g .}
We can introduce a connection $\cg (g)$
over the tangent vector bundle.  It is characterized
by its coordinate transformation property:
\eqn\etransc{\ctg (\tg) = {1 \over f'(g)} \left(
\cg (g) + {f'' (g) \over f'(g)} \right) .}
The connection can be obtained from
the operator product expansion of two
conjugate fields, which is given by
\eqn\eope{
\O_g (r) \O_g (0) \simeq {1 \over 2 \pi^2} \left(
(C_g)_g^{~{\bf 1}} (r;g) {\bf 1} + (C_g)_g^{~g} (r;g) \O_g (0)
\right) + O\left( {1 \over r^2} \right) ,}
where the composite fields of dimension higher than
four are suppressed.
The coefficient in front of the conjugate field is expressed
as the second order covariant derivative of
the beta function \rsonoda:
\eqn\ecoeff{(C_g)_g^{~g} (1;g) = - {d \over dg} \left( \left(
{d \over dg} - \cg (g) \right) \beta^g (g) \right) .}
This relation determines the connection uniquely.
For small $g$, the connection behaves as
\ref\rsuzarkesh{W.-C.~Su and A.~Zarkesh,
work in progress}
\eqn\esmallg{\cg (g) = {1 \over g} + O(1) .}
(See ref.~\ref\rsonodaII{H.~Sonoda and W.-C.~Su, ``Operator
product expansions in the two-dimensional O(N) non-linear
sigma model,'' hep-th/9406007}
for an explanation of the singular term
$1/g$ in an analogous model.)

We now define a constant $K$ by
\eqn\eK{K \equiv \lim_{g \to 0} \left( \cg (g) - {1 \over g}\right) .}
(To compute $K$, we must calculate the OPE coefficient
\ecoeff\ up to two loop order.  It has not been done
yet in any scheme.)  We define $\tilde{K}$ for the coordinate $\tg$
analogously.
The transformation property \etransc\
implies
\eqn\etransK{\tilde{K} - K = {a \over 2} ~,}
where $a$ is given by \ea.  This implies that
$\L (g)$, redefined by
\eqn\elambdatrue{\L (g) \equiv \e^{- {2 \over b_1}
\left( {1 \over g} + K \right)} g^{- {2 b_2\over 3 b_1^2}}
\exp \left[
\int_0^g dx \left( {1 \over \beta^g (x)} - {2 \over b_1 x^2}
+ {2 b_2 \over 3 b_1^2 x} \right) \right] ,}
is invariant under the change of
coordinate \escheme.  Therefore, $\L (g)$ is scheme independent.

Let $M(g)$ be the mass gap in QCD.  From the
RG, it must be proportional to $\L (g)$.  Since $M(g)$,
being physical, is independent of the scheme,
the proportionality constant must be also scheme independent:
\eqn\emass{M(g) = ({\rm universal~constant}) \cdot \L (g) .}

In this short note we have shown how to define
a scheme independent scale parameter $\L$
in terms of the beta function $\beta^g$ and connection
$\cg$.  The scheme dependence of the two quantities gets
canceled in $\L$.

\listrefs
\parindent=20pt
\bye